\documentclass[conference,letterpaper]{IEEEtran}
\IEEEoverridecommandlockouts
\usepackage{cite}
\usepackage{algorithmic}
\usepackage{textcomp}
\usepackage{xcolor}
\usepackage{soul}
\usepackage{graphicx,color}
\usepackage{amsfonts,amsthm,amssymb}
\usepackage{mathrsfs,amsmath,arydshln,latexsym}
\usepackage{cite,url}
\usepackage[bookmarks,colorlinks]{hyperref}
\usepackage[linesnumbered,ruled,vlined]{algorithm2e}
\usepackage{multirow,array,makecell}
\usepackage{stfloats}
\usepackage{fancyhdr}
\usepackage{bm}
\usepackage{enumitem}
\usepackage{pifont}
\usepackage{subcaption}
\usepackage{nomencl}
\usepackage{colortbl}
\usepackage{diagbox}
\usepackage{balance}
\usepackage{braket}
\usepackage{physics}
\usepackage{framed}
\usepackage{threeparttablex}
\allowdisplaybreaks[4]

\begin{document}

	\title{
		Rydberg Atomic Quantum Receivers for\\ the Multi-User MIMO Uplink
	}

	\author{Tierui Gong, 
		Chau Yuen, 
		Chong Meng Samson See, 
		Mérouane Debbah, 
		Lajos Hanzo 
		\vspace{-1.3cm}
		\thanks{T. Gong and C. Yuen are with School of Electrical and Electronics Engineering, Nanyang Technological University, Singapore 639798 (e-mail: trgTerry1113@gmail.com, chau.yuen@ntu.edu.sg). 
			C. M. S. See is with DSO National Laboratories, Singapore 118225 (e-mail: schongme@dso.org.sg). 
			M. Debbah is with  KU 6G Research Center, Department of Computer and Information Engineering, Khalifa University, Abu Dhabi 127788, UAE (email: merouane.debbah@ku.ac.ae) and also with CentraleSupelec, University Paris-Saclay, 91192 Gif-sur-Yvette, France.
			L. Hanzo is with School of Electronics and Computer Science, University of Southampton, SO17 1BJ Southampton, U.K. (e-mail: lh@ecs.soton.ac.uk).}
		\thanks{This work was supported by the Ministry of Education (MOE), Singapore, under its MOE Academic Research Fund (AcRF) Tier 1 Thematic Grant 023780-00001. 
		L. Hanzo would like to acknowledge the financial support of the Engineering and Physical Sciences Research Council (EPSRC) projects under grant EP/Y037243/1, EP/W016605/1, EP/X01228X/1, EP/Y026721/1, EP/W032635/1, EP/Y037243/1 and EP/X04047X/1 as well as of the European Research Council's Advanced Fellow Grant QuantCom (Grant No. 789028).}
		\vspace{-0.5cm}
	}
	
	\maketitle
	
	\pagestyle{empty}
	\thispagestyle{empty}

	\begin{abstract}
		Rydberg atomic quantum receivers exhibit great potential in assisting classical wireless communications due to their outstanding advantages in detecting radio frequency signals. To realize this potential, we integrate a Rydberg atomic quantum receiver into a classical multi-user multiple-input multiple-output (MIMO) scheme to form a multi-user Rydberg atomic quantum MIMO (RAQ-MIMO) system for the uplink. 
		To study this system, we first construct an equivalent baseband signal model, which facilitates convenient system design, signal processing and optimizations. We then study the ergodic achievable rates under both the maximum ratio combining (MRC) and zero-forcing (ZF) schemes by deriving their tight lower bounds. We next compare the ergodic achievable rates of the RAQ-MIMO and the conventional massive MIMO schemes by offering a closed-form expression for the difference of their ergodic achievable rates, which allows us to directly compare the two systems. Our results show that RAQ-MIMO allows the average transmit power of users to be $> 25$ dBm lower than that of the conventional massive MIMO.  
		Viewed from a different perspective, an extra $\sim 8.8$ bits/s/Hz/user rate becomes achievable by ZF RAQ-MIMO. 
	\end{abstract}


	\section{Introduction}
	
	In the era of the second quantum revolution, quantum science rests on three main pillars, namely, quantum computing, quantum communications, and quantum sensing \cite{dowling2003quantum}. Specifically, quantum sensing relies on quantum phenomena to realize the measurements of a physical quantity at an unprecedented accuracy \cite{degen2017quantum}. At the time of writing, physical quantities can be precisely measured is the electric field, magnetic field, frequency, temperature, pressure, rotation, acceleration, and so forth \cite{gschwendtner2024quantum}. The Rydberg atomic quantum receiver (RAQR) constitutes one of the emerging quantum sensing technologies, which is capable of detecting the classical radio frequency (RF) signals' amplitude, phase, polarization, and direction-of-arrival (DOA) etc., at an unparalleled precision \cite{gong2024RAQRs,gong2025rydberg}. 
	
	RAQRs have numerous compelling features, including ultra-high sensitivity, broadband tunability, compact form factor, and International System of Units (SI)-traceability \cite{gong2024RAQRs,Fancher2021Rydberg}. Specifically, benefiting from the large dipole moments of Rydberg atoms, the sensitivity has been experimentally improved to nV/cm/$\sqrt{\text{Hz}}$ \cite{jing2020atomic,borowka2024continuous} and could potentially reach pV/cm/$\sqrt{\text{Hz}}$. Additionally, the broadband tunability of RAQRs reveals that they are capable of receiving RF signals distributed over a wide frequency range (from near direct-current to Terahertz) using only a single vapor cell. RAQRs also have a compact form factor, because the size is independent of the wavelength of RF signals. Furthermore, they directly downconvert passband RF signals to baseband without requiring any complex integrated circuits (IC). Last but not least, the measurements carried out by RAQRs are directly linked to the SI constant without requiring any calibration. 
	
	These distinctive features of RAQRs have the promise of revolutionizing conventional antenna based RF receivers, because they outperform conventional antenna based RF receivers without requiring sophisticated antenna calibration \cite{holloway2014broadband}, especially for realizing massive multiple-input multiple-output (M-MIMO) \cite{Larsson2014Massive} and holographic MIMO \cite{Gong2023Holographic, Gong2024HMIMO, Gong2024Near} systems. Hence, RAQRs have the potential of supporting the extreme data rate, connectivity, and latency specifications of next-generation wireless systems \cite{recommendation2023framework}.

	To fully unlock the potential of RAQRs, initial experimental studies have been performed for verifying the capabilities, including the detection of amplitudes \cite{Fancher2021Rydberg}, phase \cite{jing2020atomic}, polarization \cite{sedlacek2013atom}, modulation \cite{nowosielski2024warm}, spatial direction \cite{robinson2021determining}, spatial displacement \cite{zhang2023quantum}, and the experimental verifications of RAQRs in assisting classical communications \cite{yuan2023rydberg}. 
	Additionally, a comprehensive overview of RAQRs harnessed for classical wireless communications and sensing was presented in \cite{gong2024RAQRs}. The construction of a fundamental signal model for an RAQR-aided wireless single-input single-output (SISO) for communications and sensing systems was presented in \cite{gong2024rydberg}. By contrast, in this paper, we consider multi-user uplink receive arrays formed by RAQRs. The resultant RAQ-MIMO system is more complex, but more inline with state-of-the-art communication scenarios. We first construct an equivalent baseband signal model for RAQ-MIMOs and then analyse the ergodic achievable rate under both the maximum ratio combining (MRC) and zero-forcing (ZF) schemes. We also critically appraise both RAQ-MIMO schemes and show their superiority over their conventional M-MIMO counterparts.

	\textit{Organization and Notations}: 
	The article is organized as follows. In Section \ref{sec:SysModel}, we outline the system model of a receiver array formed by superheterodyne RAQRs. In Section \ref{sec:BBSigModel}, we then detail the equivalent baseband signal model. In Section \ref{sec:EAR}, we analyse the ergodic achievable rate and perform detailed comparisons. We present our simulation results in Section \ref{sec:Simulations}, and finally conclude in Section \ref{sec:Conclusions}. 
	The notations, $\jmath^2 = 1$, 
	$\mathscr{R} \{ \cdot \}$ and $\mathscr{I} \{ \cdot \}$ represent the real and imaginary parts of a complex number; $\chi'$ represents the derivative of $\chi$; $\hbar$ denotes the reduced Planck constant; $c$ and $\epsilon_0$ are the speed of light in free space and the vacuum permittivity. $k_B$ and $q$ are the Boltzmann constant and elementary charge, respectively. 

	\begin{figure*}[t!]
		\centering
		\includegraphics[height=3.6cm, width=16.6cm]{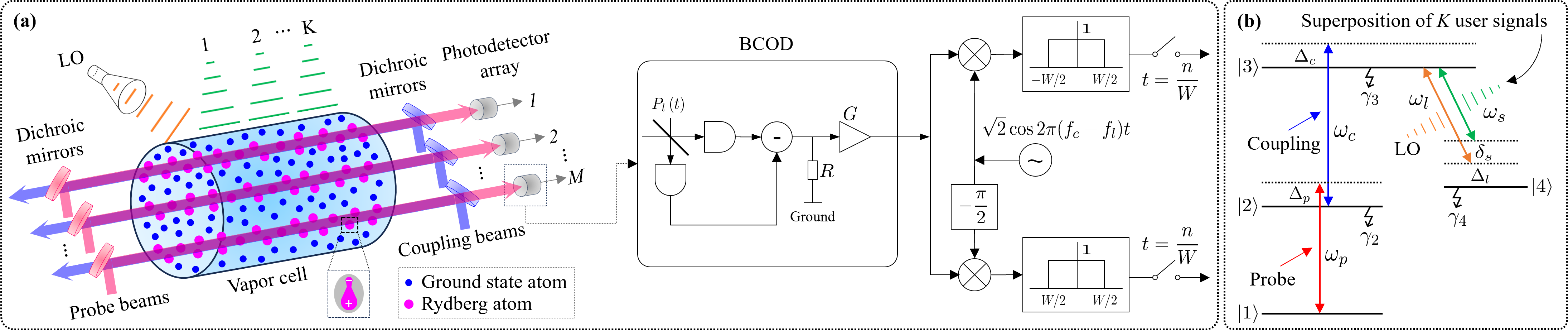}
		\caption{(a) The superheterodyne scheme of RAQ-MIMO, and (b) its corresponding four-level scheme.}
		\vspace{-1.2em}
		\label{fig:RAQMIMOScheme}
	\end{figure*}

	\section{System Model of RAQ-MIMO} 
	\label{sec:SysModel}
	
	For exploiting the potential ultra-high sensitivity in both amplitude and phase extraction, we apply the superheterodyne philosophy for our RAQ-MIMO receiver. Its implementation is highlighted in Fig \ref{fig:RAQMIMOScheme}(a), where a vapor cell containing Cesium (Cs) atoms is spatially penetrated by multiple pairs of counter-propagating laser beams (probe and coupling) to prepare the Rydberg atoms, which relies on $M$ receive elements. 
	Our RAQ-MIMO receiver has a local oscillator (LO) having a resonant (or near-resonant) frequency, which means that the carrier frequency of the LO is exactly the same (or has a small frequency difference) as the on-resonance frequency between the two Rydberg states. 
	The multiple RF signals of $K$ simultaneous users having the same carrier frequency are received by the RAQ-MIMO. 
	The superposition of the LO and the $K$ RF user signals will affect Rydberg atoms by stimulating electrons transitions between two Rydberg states.  This will subsequently influence the probe beams that are optically read-out by a photodetector array. 
	The Rydberg atomic responses transform the RF signals to the optical domain. 
	
	The Rydberg atomic response of each receive element of the RAQ-MIMO can be depicted by the four-level scheme, seen in Fig. \ref{fig:RAQMIMOScheme}(b). Briefly, four energy levels, namely the ground state $\ket{1}$, immediate state $\ket{2}$, and Rydberg states $\ket{3}$ and $\ket{4}$, are involved in this scheme. They are coupled by the probe beam, coupling beam, and the superposition of RF signals, resonant or near-resonant, respectively. The probe, coupling, and superimposed RF signals have their own so-called Rabi frequencies $\Omega_{p,m}$, $\Omega_{c,m}$, and $\Omega_{\text{RF},m}$, respectively, for the $m$-th receive element, which correspond to their own signal powers. We assume that the powers of the probe and coupling beams are identical for all $M$ receive elements, implying $\Omega_{p,m} = \Omega_p$ and $\Omega_{c,m} = \Omega_c$ for $m = 1, \cdots, M$. By contrast, the powers of the superimposed RF signals are different for all $M$ receive elements.
	More particularly, $\Omega_{\text{RF},m}$ can be approximated to a linear combination form 
	$\Omega_{\text{RF}, m} \approx \Omega_{l} + \sum_{k=1}^{K} \Omega_k \cos{(2 \pi f_{\delta} t + \theta_{\delta, k,m})}$, where $\Omega_{l}$ and $\Omega_k$ denote the Rabi frequencies of the LO and of the $k$-th user, $f_{\delta} = f_c - f_l$ and $\theta_{\delta,k,m} = \theta_{x,k} - \theta_{y_m}$ represent the frequency difference between the RF user signals and the LO, and the phase difference between the $k$-th user and the LO. We assume $\Omega_{k} \ll \Omega_{l}$, $k = 1, \cdots, K$, implying that the impinging RF signals are weak compared to the LO. This assumption is reasonable and realizable in realistic implementations \cite{jing2020atomic}.
	
	The detuning frequency represents a small frequency difference between the required coupling frequency for the electron transition of two energy levels and the actual frequency of the external electromagnetic wave to enable this transition. 
	Specifically, the detuning frequencies corresponding to the probe beam, the coupling beam, and the LO, seen in Fig. \ref{fig:RAQMIMOScheme}(b), are $\Delta_p = \omega_{12} - \omega_p$, $\Delta_c = \omega_{23} - \omega_c$, and $\Delta_l = \omega_{34} - \omega_l$, respectively, where $\omega_{i-1,i}$ is the on-resonance angular frequency between level $\ket{i-1}$ and $\ket{i}$, $i=2, 3, 4$, and $\omega_{p, c, l}$ represent the angular frequency of the probe beam, the coupling beam, and the LO, respectively. When the laser beams and the LO are resonant, their detuning frequencies are zero. 
	Let us denote the spontaneous decay rate of the $i$-th level by $\gamma_{i}$, where $i = 2, 3, 4$, the relaxation rates related to the atomic transition effect and collision effect by $\gamma$ and $\gamma_c$, respectively. For simplicity, we assume $\gamma = \gamma_c = 0$. Additionally, we assume that the decay rates of level $\ket{3}$ and level $\ket{4}$ are comparatively small so that they can be reasonably ignored, i.e., $\gamma_3 = \gamma_4 = 0$.

	\subsection{RF-to-Optical Transformation Model of RAQ-MIMO} 
	
	Let us denote the amplitude, frequency, and initial phase of the $m$-th probe beam at the access area of the vapor cell, respectively, by $U_{0,m}$, $f_{p}$, and $\phi_{0,m}$. After propagating through the vapor cell, the probe beam is affected by the Rydberg atoms in terms of its amplitude and phase at the output area of the vapor cell. 
	Upon denoting the amplitude and the phase of the $m$-th output probe beam, respectively, by $U_{p, m} (\Omega_{\text{RF},m})$ and $\phi_{p, m} (\Omega_{\text{RF},m})$, they can be associated with their input counterparts following \cite{gong2024rydberg}
	\begin{align}
		\label{eq:AmplitudeRelation}
		U_{p, m} (\Omega_{\text{RF},m}) 
		&= U_{0,m} \exp \left( - \frac{\pi L}{\lambda_p} \mathscr{I} \{\chi_{m} (\Omega_{\text{RF},m}) \} \right), \\
		\label{eq:PhaseRelation}
		\phi_{p, m} (\Omega_{\text{RF},m})  
		&= \phi_{0,m} + \frac{\pi L}{\lambda_p} \mathscr{R} \{\chi_{m} (\Omega_{\text{RF},m}) \}, 
	\end{align}
	where $\lambda_p$ is the wavelength of the probe laser, $L$ denotes the length of the vapor cell, and $\chi_{m} (\Omega_{\text{RF},m})$ is the susceptibility of the atomic vapor medium. 
	Furthermore, $\chi_{m} (\Omega_{\text{RF},m})$ in \eqref{eq:AmplitudeRelation} and \eqref{eq:PhaseRelation} was formulated as \cite{gong2024rydberg}
	\begin{align}
		\nonumber
		\chi_{m} (\Omega_{\text{RF},m}) 
		&= - D \frac{{{A_{1,m}}\Omega _{{\rm{RF}},m}^4 + {A_{2,m}}\Omega _{{\rm{RF}},m}^2 + {A_{3,m}}}}{{{C_{1,m}}\Omega _{{\rm{RF}},m}^4 + {C_{2,m}}\Omega _{{\rm{RF}},m}^2 + {C_{3,m}}}} \\
		\label{eq:Susceptibility}
		&+ \jmath D \frac{{{B_{1,m}}\Omega _{{\rm{RF}},m}^4 + {B_{2,m}}\Omega _{{\rm{RF}},m}^2 + {B_{3,m}}}}{{{C_{1,m}}\Omega _{{\rm{RF}},m}^4 + {C_{2,m}}\Omega _{{\rm{RF}},m}^2 + {C_{3,m}}}}, 
	\end{align}
	where we have $D = \frac{2 N_0 \mu_{12}^{2}}{\epsilon_0 \hbar}$, $N_0$ is the atomic density in the vapor cell, and $\mu_{12}$ is the dipole moment of the transition $\ket{1} \rightarrow \ket{2}$. The coefficients $A_{1,m}, A_{2,m}, A_{3,m}$, $B_{1,m}, B_{2,m}, B_{3,m}$, $C_{1,m}, C_{2,m}, C_{3,m}$ are presented in (45)-(53) of the Appendix A of \cite{gong2024rydberg}. 
	For further use, the real and imaginary parts of the derivative of $\chi_{m} (\Omega_{\text{RF},m})$ are \cite{gong2024rydberg} 
	\begin{align}
		\nonumber
		&\mathscr{R} \{ \chi'_{m} \left( \Omega_{l} \right) \} 
		= - 2 D {\Omega_{l}} \left[ \frac{{  {2{A_{1,m}}\Omega_{l}^2 + {A_{2,m}}} }}{{{C_{1,m}}\Omega _{l}^4 + {C_{2,m}}\Omega _{l}^2 + {C_{3,m}}}} \right. \\
		\nonumber
		&\left. - \frac{{ \left( {{A_{1,m}}\Omega _{l}^4 + {A_{2,m}}\Omega _{l}^2 + {A_{3,m}}} \right) \left( {2{C_{1,m}}\Omega_{l}^2 + {C_{2,m}}} \right)}}{{{{\left( {{C_{1,m}}\Omega_{l}^4 + {C_{2,m}}\Omega _{l}^2 + {C_{3,m}}} \right)}^2}}} \right], \\
		\nonumber
		&\mathscr{I} \{ {\chi'_{m}} \left( \Omega_{l} \right) \}
		= 2 D {\Omega_{l}} \left[ \frac{{  {2{B_{1,m}}\Omega_{l}^2 + {B_{2,m}}} }}{{{C_{1,m}}\Omega_{l}^4 + {C_{2,m}}\Omega _{l}^2 + {C_{3,m}}}} \right. \\
		\nonumber
		&\left. - \frac{{ \left( {{B_{1,m}}\Omega _{l}^4 + {B_{2,m}}\Omega _{l}^2 + {B_{3,m}}} \right) \left( {2{C_{1,m}}\Omega_{l}^2 + {C_{2,m}}} \right)}}{{{{\left( {{C_{1,m}}\Omega _{l}^4 + {C_{2,m}}\Omega _{l}^2 + {C_{3,m}}} \right)}^2}}} \right]. 
	\end{align}

	Based on \eqref{eq:AmplitudeRelation} and \eqref{eq:PhaseRelation}, the probe beam at the output of the vapor cell for the $m$-th receive element can be formulated as
	\begin{align}
		\nonumber
		P_{m} (\Omega_{\text{RF},m}, t) 
		&= \bar{U}_{p,m} (\Omega_{\text{RF},m}) \cos \left( 2 \pi f_{p} t + \phi_{p,m} (\Omega_{\text{RF},m}) \right) \\
		\label{eq:OutputProbeSig}
		&= \sqrt{2} \mathscr{R} \left\{ P_{b,m} (\Omega_{\text{RF},m}, t) e^{\jmath 2 \pi f_p t} \right\}.
	\end{align}
	In \eqref{eq:OutputProbeSig}, $\bar{U}_{p,m} \triangleq \sqrt{ \frac{\pi c \epsilon_0}{\ln {2}} } \frac{F_p}{2} U_{p,m}$ is the equivalent amplitude, where $F_{p}$ is the full width at half maximum (FWHM) of the probe beam. $P_{b,m} (\Omega_{\text{RF}}, t) \triangleq \sqrt{\mathcal{P}_{m} (\Omega_{\text{RF},m})} e^{\jmath \phi_p (\Omega_{\text{RF},m})}$ is the equivalent baseband signal of the output (passband) probe beam, which has an average power of $\mathcal{P}_{m} (\Omega_{\text{RF},m}) = \frac{\pi c \epsilon_0}{8 \ln {2}}  F_p^2 \left| U_{p,m}(\Omega_{\text{RF},m}) \right|^{2}$. 
	The probe signal is detected by a photodetector, which exploits a balanced coherent optical detection (BCOD) scheme. Therein, the probe beam and a local optical beam are combined to form two distinct optical sources, namely ${P_1} = \frac{1}{{\sqrt 2 }}\left[ {P_l}\left( t \right) - P_{m} (\Omega_{\text{RF},m}, t) \right]$ and ${P_2} = \frac{1}{{\sqrt 2 }}\left[ {P_l}\left( t \right) + P_{m} (\Omega_{\text{RF},m}, t) \right]$, which are then detected by two photodetectors, respectively. The consequent photocurrents are subtracted to produce a voltage across a resistance load (assume unit ohm). A low-noise amplifier (LNA) having a gain of $G$ is applied to obtain the output of the BCOD scheme. This output is down-converted to a baseband signal by removing the intermediate frequency $f_c - f_l$, as shown in Fig. \ref{fig:RAQMIMOScheme}.

	\subsection{Equivalent Baseband Signal Model of RAQ-MIMO}
	\label{sec:BBSigModel}
	
	Given the equivalent baseband signal model of an RAQR-SISO system in \cite{gong2024rydberg}, we extend it to the RAQ-MIMO case, where multiple receive elements simultaneously detect multiple RF signals impinging from different users, as demonstrated in Fig \ref{fig:RAQMIMOScheme}(a). The equivalent baseband signal model of the $m$-th receive element can be formulated as 
	\begin{align}
		\label{eq:NarrowbandSampledOutputPlusNoise}
		\tilde{V}_{m} = \sqrt{A_e {\varrho_m}} \Phi_m \sum_{k=1}^{K} h_{m,k} s_{k} + w_m,
	\end{align}
	where $A_e$ is the $m$-th effective receiver aperture; $\varrho_m$ and $\Phi_m$ are the gain and phase shift of the $m$-th receive element of RAQ-MIMO, respectively; $h_{m,k}$ is the wireless channel between the $k$-th user and the $m$-th receive element; $s_{k}$ is the transmit signal of the $k$-th user; and $w_m$ is the noise contaminating the signal of the $m$-th receive element. 

	In this system, we assume that all the $M$ receiver apertures are identical.  
	Note that the effective receiver aperture of the RAQ-MIMO are independent of the RF wavelength. 
	In the BCOD scheme, let us denote the power and phase of the local optical signal by $\mathcal{P}_{m}^{(l)}$ and ${\phi_{m}^{(l)}}$ for the $m$-th channel, respectively. Then $\varrho_{m}$ and $\Phi_{m}$ in \eqref{eq:NarrowbandSampledOutputPlusNoise} are given by \cite{gong2024rydberg}
	\begin{align}
		\label{eq:GainBCOD}
		\sqrt{\varrho_m} &= 2 K_{\text{gain}} \sqrt{\mathcal{P}_{m}^{(l)} \mathcal{P}_{m} (\Omega_{l})} {\kappa_{m}}({\Omega _l}), \\
		\label{eq:PhaseBCOD}
		\Phi_{m} &= \frac{1}{2} \left[ e^{ - \jmath \left( {\theta_{y_m}} - \varphi_{m} ({\Omega _l}) \right) } + e^{ - \jmath \left( {\theta_{y_m}} + \varphi_{m} ({\Omega _l}) \right) } \right], 
	\end{align}
	where we have ${K_{\text{gain}} \triangleq \frac{{\sqrt{G} \eta q}}{{\hbar \omega_p \sqrt{c \epsilon_0 A_e} }}}$,  
	\begin{align}
		\label{eq:Kappa2}
		\kappa_{m} (\Omega_l) 
		&= \frac{\pi L \mu_{34}}{\hbar \lambda_p} \sqrt {{{\left[ {{\mathscr I}\{ \chi'_{m} ({\Omega _l})\} } \right]}^2} + {{\left[ {{\mathscr R}\{ \chi'_{m} ({\Omega _l})\} } \right]}^2}}, \\
		\label{eq:varphi_2}
		{\varphi_{m} ({\Omega _l})} 
		&= {\phi_{m}^{(l)}} - {\phi_{p, m}}({\Omega _l}) + {\psi_{p, m}}({\Omega _l}). 
	\end{align}
	In \eqref{eq:varphi_2}, $\psi_{p, m} (\Omega_{l}) = \arccos \frac{ {{\mathscr I}\{ \chi'_{m} ({\Omega _l})\} } }{ \sqrt {{{\left[ {{\mathscr I}\{ \chi'_{m} ({\Omega _l})\} } \right]}^2} + {{\left[ {{\mathscr R}\{ \chi'_{m} ({\Omega _l})\} } \right]}^2}} }$. 
	
	Upon expressing \eqref{eq:NarrowbandSampledOutputPlusNoise} in matrix form by collecting all $M$ measurements, we arrive at 
	\begin{align}
		\label{eq:NarrowbandSampledOutputPlusNoise_MatrixForm}
		\bm{v} = \sqrt{A_e} \bm{\varTheta} \bm{H} \bm{s} + \bm{w},
	\end{align}
	where $\bm{\varTheta} = \text{diag} \{ \sqrt{\varrho_1} \Phi_1, \sqrt{\varrho_2} \Phi_2, \cdots, \sqrt{\varrho_M} \Phi_M \}$, $\bm{H}$ is the channel matrix, $\bm{s}$ is the transmit data vector, and $\bm{w}$ is the noise vector. Let us consider an uncorrelated Rayleigh fading channel $\bm{H} = \bm{G} \bm{\Sigma}$, where $\bm{G} \in \mathbb{C}^{M \times K}$ is the small-scale fading channel matrix, which is asymptotically orthogonal for large $M$. $\bm{\Sigma} = \text{diag} \{ \sqrt{\beta_{1}}, \sqrt{\beta_{2}}, \cdots, \sqrt{\beta_{K}} \}$ represents the large-scale fading of the uplink $K$ transmitters. 
	Finally, $\bm{w}$ is the complex additive white Gaussian noise (AWGN), namely we have $\bm{w} \sim \mathcal{CN} (\bm{0}, \sigma^2 \bm{I}_{M})$. Specifically, $\sigma^2 = {\cal P}_{\text{SQL}} = 4 G \left( \frac{{\eta q}}{{\hbar \omega_p }} \right)^{2} \mathcal{P}_{m}^{(l)} \mathcal{P}_{m} (\Omega_{l}) {\kappa_{m}^{2}}({\Omega _l}) \cos^{2} {\varphi_{m} ({\Omega _l})} \left( \frac{E_{\text{SQL}}}{\sqrt{B}} \right)^{2} B$ denotes the noise power corresponding to the standard quantum limit (SQL), where $\frac{E_{\text{SQL}}}{\sqrt{B}} = \frac{\hbar}{ \mu_{34} \sqrt{N T_2}}$ \cite{Fancher2021Rydberg}. 

	To make the signal model of RAQ-MIMO more specific, 
	we assume that the $M$ receive elements form a uniform linear array (ULA), having a spacing (denoted by $d$) less than or equal to half-a-wavelength of the RF signal. 
	We assume that the LO signal is a plane wave, so that the amplitudes are approximately identical at all receive elements, but the phase difference increases uniformly compared to the reference receive element, which is expressed as 
	\begin{align}
		\theta_{y_m} = \theta_{y_1} + \frac{2 \pi}{\lambda} d (m-1) \sin \vartheta, 
	\end{align} 
	where $\theta_{y_1}$ denotes the phase of the reference (first) receive element, $\vartheta$ is the angle of the incident LO signal, and $\lambda$ denotes the wavelength of the user signals. 
	Additionally, for a well-configured RAQ-MIMO, we have ${\cal P}_{m}^{(l)} \triangleq {\cal P}_{l}$, $\phi_{m}^{(l)} \triangleq \phi_{l}$, ${\cal P}_{m} (\Omega_{l}) \triangleq {\cal P} (\Omega_{l})$, ${\kappa_{m}}({\Omega _l}) \triangleq {\kappa}({\Omega _l})$, and $\varphi_{m} (\Omega_{l}) \triangleq \varphi (\Omega_{l})$ for all $M$ receive elements. 
	Therefore, \eqref{eq:GainBCOD} and \eqref{eq:PhaseBCOD} become $\sqrt{\varrho_{m}} \triangleq \sqrt{\varrho} = 2K \sqrt{\mathcal{P}_{l} \mathcal{P} (\Omega_{l})} {\kappa}({\Omega _l})$ and $\Phi_{m} = \Phi e^{ - \jmath \frac{2 \pi}{\lambda} d (m-1) \sin \vartheta }$, where $\Phi \triangleq \Phi_{1}$. 
	Based on the above discussions, we obtain $\bm{\varTheta} = \sqrt{\varrho} \Phi \bm{D}$ and reformulate \eqref{eq:NarrowbandSampledOutputPlusNoise_MatrixForm} as 
	\begin{align}
		\label{eq:NarrowbandSampledOutputPlusNoise_MatrixForm_SC}
		\bm{v} = \sqrt{A_e \varrho} \Phi \bm{D} \bm{H} \bm{s} + \bm{w}, 
	\end{align}
	where $\bm{D} = \text{diag} \{ 1, e^{ - \jmath \frac{2 \pi}{\lambda} d \sin \vartheta }, \cdots, e^{ - \jmath \frac{2 \pi}{\lambda} d (M-1) \sin \vartheta } \}$.

	\section{Ergodic Achievable Rate of RAQ-MIMO}
	\label{sec:EAR}
	
	We assume that the RAQ-MIMO has perfect channel state information (CSI). The linear receiver combining matrix used for recovering the transmit signal based on the measurements is denoted by $\bm{C}$.
	The $k$-th user's signal estimate is given by 
	\begin{align}
		\nonumber
		r_{k} 
		&= \sqrt{A_e} \bm{c}_{k}^{*} \bm{\varTheta} \bm{h}_{k} s_{k} + \sqrt{A_e} \sum_{i=1, \ne k}^{K} \bm{c}_{k}^{*} \bm{\varTheta} \bm{h}_{i} s_{i} + \bm{c}_{k}^{*} \bm{w}. 
	\end{align}
	Upon assuming an uncorrelated fading channel $\bm{h}_{k} = \sqrt{\beta_k} \bm{g}_{k}$, where $\beta_k$ denotes the large-scale fading and $\bm{g}_{k} \sim \mathcal{CN}(\bm{0}, \bm{I}_{M})$ represents the small-scale fading, we can obtain the ergodic achievable rate of the $k$-th user as 
	\begin{align}
		\label{eq:AchievableRate_MRC}
		R_{k} 
		&= \mathbb{E} \left\{ \log_{2} \left( 1 + \gamma_{k} \right) \right\}, 
	\end{align}
	where the signal-to-interference-plus-noise ratio (SINR) is
	\begin{align}
		\label{eq:Gamma_MRC}
		\gamma_{k}
		&= \frac{ {\cal P}_{s} A_e \left| \bm{c}_{k}^{*} \bm{\varTheta} \bm{h}_{k} \right|^{2} }{ {\cal P}_{s} A_e \sum_{i=1, \ne k}^{K} \left| \bm{c}_{k}^{*} \bm{\varTheta} \bm{h}_{i} \right|^{2} + \sigma^{2} \left\| \bm{c}_{k} \right\|^{2} }. 
	\end{align}

		\subsection{The Lower Bound of the Ergodic Achievable Rate}
		
		First, we apply the MRC scheme characterized by $\bm{C} = \bm{\varTheta} \bm{H}$ for \eqref{eq:NarrowbandSampledOutputPlusNoise_MatrixForm_SC}, and provide the following lower bound (LB) result for RAQ-MIMO 
		with the aid of perfect CSI 
		\begin{align}
			\label{eq:AchievableRate_MRC_LB_SC}
			&R_{k} \ge 
			{\log _2}\left[ 1 + \frac{{ \varrho \cos^{2} \varphi (\Omega_{l}) \frac{{\cal P}_s {A_e}}{\sigma^2} (M-1) {\beta _k}}}{ 1 + {\varrho \cos^{2} \varphi (\Omega_{l})} \frac{{\cal P}_s {A_e}}{\sigma^2}\sum_{i = 1, \ne k}^K {{\beta _i}} } \right]. 
		\end{align}

		We then apply the ZF scheme characterized by $\bm{C} = \bm{\varTheta} \bm{H} \left( \bm{H}^{*} \bm{\varTheta}^{*} \bm{\varTheta} \bm{H} \right)^{-1}$ for \eqref{eq:NarrowbandSampledOutputPlusNoise_MatrixForm_SC} and proceed to determine the corresponding rate LB. Explicitly, under the assumption of 
		perfect CSI and the ZF criterion, the LB of is given by
		\begin{align}
			\label{eq:AchievableRate_ZF_LB_SC}
			&R_{k} \ge 
			{\log _2}\left[ 1 + \frac{ {\varrho}{{\cos }^2}\varphi ({\Omega _l}) {{\cal P}_s}{A_e} \left( {M - K} \right) {\beta _k} }{{{\sigma ^2}}} \right]. 
		\end{align}
		See Appendix \ref{Appendix:LBs} for the proofs of \eqref{eq:AchievableRate_MRC_LB_SC} and \eqref{eq:AchievableRate_ZF_LB_SC}.

		Before comparing the proposed RAQ-MIMO to the conventional M-MIMO system, we present the LB of the achievable rate of conventional M-MIMO based on \cite{ngo2013energy} as follows
		\begin{align}
			\label{eq:MIMO_AchievableRate_MRC_LB}
			&\tilde{R}_{k} \ge 
			\begin{cases}
				{\log _2}\left( {1 + \frac{{ \varrho_{0} {{\cal P}_s}{A_0} (M-1) {\beta _k}}}{{ {\varrho_{0}} {{\cal P}_s}{A_0}\sum_{i = 1, \ne k}^K {{\beta _i}} + {\sigma_{\text{m-mimo}} ^2} }}} \right) & \text{MRC}, \\
				{\log _2}\left( {1 + \frac{ \varrho _0 {{\cal P}_s{A_0}\left( {M - K} \right) {\beta _k}}}{{{\sigma_{\text{m-mimo}} ^2}}}} \right) & \text{ZF},
			\end{cases} 
		\end{align}
		where $\varrho_{0}$ is the gain of each RF chain of M-MIMO and $A_0$ is the effective receiver aperture of an M-MIMO antenna. In \eqref{eq:MIMO_AchievableRate_MRC_LB} $\varrho_{0} \triangleq G_{\text{Ant}} G_{\text{LNA}}$ is determined by the antenna gain $G_{\text{Ant}}$ and the LNA gain $G_{\text{LNA}}$ employed for each RF chain. 
		We assume that each antenna of the array is a half-wavelength dipole antenna. Then the effective receiver aperture of each antenna is $A_0 = \lambda^2 / (4 \pi)$.
		The noise power of M-MIMO is different from that of the RAQ-MIMO. Let us assume that identical electronic components are applied for the RAQ-MIMO and M-MIMO, so that they face an identical LNA noise. Therefore, we have $\sigma_{\text{m-mimo}}^{2} = {\cal P}_{\text{th}} + {\cal P}_{\text{LNA}} = G_{\text{LNA}} k_B T_{\text{room}} B + k_B T_{\text{LNA}} B$.

		We present the following cases for gaining deeper insights.
		\begin{enumerate}[label={\em {C\arabic*}}, leftmargin=1.7em, labelindent=0pt, itemindent = 0em]
			\item \label{itm:C1} 
			For the MRC receivers of both RAQ-MIMO and M-MIMO, the LBs of $R_{k}$ and $\tilde{R}_{k}$ become identical if the inter-user interferences dominate the noise terms. The result is given by ${\log _2}\left( 1 + \frac{ (M-1) {\beta _k}}{ \sum_{i = 1, \ne k}^K {{\beta _i}} } \right)$. It reveals that the ergodic achievable rates become saturated for a given $M$ and become identical for MRC RAQ-MIMO and MRC M-MIMO in the face of high interferences. 

			\item \label{itm:C3} 
			For both MRC and ZF, we have $R_{k} \rightarrow$ 
			${\log _2}\left( {1 + \frac{{ {\cal E} {\beta _k} {A_e} {\varrho}{{\cos }^2}\varphi ({\Omega _l})}}{{{\sigma ^2}}}} \right)$, $\tilde{R}_{k} \rightarrow$ 
			${\log _2}\left( {1 + \frac{{ {\cal E} {\beta _k} {A_0} {\varrho_{0}}}}{{\sigma_{\text{m-mimo}}^{2}}}} \right)$ for large $M$ if the transmit power of each user scales with $M$ via ${\cal P}_{s} = \frac{{\cal E}}{M}$ for a fixed ${\cal E}$.
		\end{enumerate}

		\subsection{Comparisons of the Ergodic Achievable Rates}
		
		For comparing the ZF and MRC receivers of RAQ-MIMO, we focus on the ratio of the SINRs in the log expressions of \eqref{eq:AchievableRate_ZF_LB_SC} and of \ref{itm:C1} above. 
		The ratio is obtained as follows. 
		\begin{align}
			\nonumber
			\mathsf{Ratio}_{1} 
			&= \left( \frac{ M - K }{ M-1 } \right) \frac{ {\varrho} {{\cos }^2} \varphi ({\Omega _l}) {{\cal P}_s} {A_e} \sum_{i = 1, \ne k}^K {{\beta _i}} }{{\sigma ^2}} \\ 
			\label{eq:SINRratio1}
			&= {\varrho} \left( \frac{ M - K }{ M-1 } \right) \frac{ {{\cal P}_{r}^{\text{(tot)}}} - {{\cal P}_{r}^{(k)}} }{{\sigma ^2}}, \\
			\label{eq:SINRratio1Asmp}
			&\overset{M \rightarrow \infty}{\longrightarrow}
			{\varrho} \frac{ {{\cal P}_{r}^{\text{(tot)}}} - {{\cal P}_{r}^{(k)}} }{{\sigma ^2}}, 
		\end{align}
		where ${{\cal P}_{r}^{\text{(tot)}}} \triangleq {{\cal P}_s} {A_e} \sum_{i = 1}^K {{\beta _i}}$ denotes the total receive power of all $K$ users at the RAQ-MIMO and ${{\cal P}_{r}^{(k)}} \triangleq {{\cal P}_s} {A_e} {\beta _k}$ is the receive power of the $k$-th user. 
		In \eqref{eq:SINRratio1}, we get the result by configuring ${{\cos }^2}\varphi ({\Omega _l}) = 1$ that is realized by setting \eqref{eq:varphi_2} to zero through ${\phi_{m}^{(l)}} = {\phi_{m}}({\Omega _l}) - {\psi_{m}}({\Omega _l})$. This configuration maximizes $\mathsf{Ratio}_{1}$. Then \eqref{eq:SINRratio1Asmp} is obtained for large $M$.
		In the high SINR regime, $\mathsf{Ratio}_{1}$ has an extra achievable rate contribution for ZF over MRC, given by $\Delta R_{k} = \log_{2} \left( \mathsf{Ratio}_{1} \right)$
		\begin{align}
			\label{eq:SE1}
			\Delta R_{k} 
			&= \log_{2} \left( \varrho \right) 
			+ \log_{2} \frac{ {{\cal P}_{r}^{\text{(tot)}}} - {{\cal P}_{r}^{(k)}} }{{\sigma ^2}} + \Delta R \\
			\label{eq:SE1Asmp}
			&\overset{M \rightarrow \infty}{\longrightarrow}
			\log_{2} \left( \varrho \right) 
			+ \log_{2} \frac{ {{\cal P}_{r}^{\text{(tot)}}} - {{\cal P}_{r}^{(k)}} }{{\sigma ^2}}, 
		\end{align}
		where we have $\Delta R \triangleq \log_{2} \left( 1 - \frac{ K - 1 }{ M-1 } \right)$. 

		For comparing the ZF of RAQ-MIMO and M-MIMO, we similarly focus our attention on the ratio of the SINRs in the log expressions of \eqref{eq:AchievableRate_ZF_LB_SC} and of \eqref{eq:MIMO_AchievableRate_MRC_LB}. 
		The ratio is obtained as 
		\begin{align}
			\label{eq:SINRratio2}
			\mathsf{Ratio}_{2} 
			&= \frac{ {\varrho} }{{\varrho _0}} \times {\frac{{{A_e}}}{{{A_0}}}} \times {\frac{{\sigma _{{\text{m-mimo}}}^2}}{{{\sigma ^2}}}},
		\end{align}
		where we set ${{\cos }^2}\varphi ({\Omega _l}) = 1$ by configuring ${\phi_{m}^{(l)}} = {\phi_{m}}({\Omega _l}) - {\psi_{m}}({\Omega _l})$.  In the high SINR regime, this ratio results in an extra achievable rate contribution for RAQ-MIMO over M-MIMO given by $\Delta \tilde{R}_{k} = \log_{2} \left( \mathsf{Ratio}_{2} \right)$, which is expressed as
		\begin{align}
			\label{eq:SE2Asmp}
			\Delta \tilde{R}_{k} 
			&= \log_{2} \frac{ \varrho }{ \varrho_0 } 
			+ \log_{2} \frac{ A_e }{ A_0 } 
			- \log_{2} \frac{ \sigma^{2} }{ \sigma_{\text{m-mimo}}^{2} }.
		\end{align}

		\section{Simulation Results}
		\label{sec:Simulations}
		To characterize the performance of RAQ-MIMO and verify its potential, in this section we present simulations quantifying its average achievable rate versus (vs.) diverse parameters.

		\subsection{Simulation Configurations}
		
		In the following simulations, we use a vapor cell having a length of $L = 10$ cm filled with Cs atoms at an atomic density of $N_0 = 3 \times 10^{10}$ $\text{cm}^{-3}$, where the population rate is 0.1\%.
		The four-level transition system of \ref{fig:RAQMIMOScheme} is 6S\textsubscript{\scalebox{0.8}{1/2}} $\rightarrow$ 6P\textsubscript{\scalebox{0.8}{3/2}} $\rightarrow$ 47D\textsubscript{\scalebox{0.8}{5/2}} $\rightarrow$ 48P\textsubscript{\scalebox{0.8}{3/2}}. The parameters of the probe beam are: wavelength of $852$ nm, beam diameter of $1.7$ mm, power of $29.8$ \textmu W, and Rabi frequency of $2 \pi \times 5.7$ MHz. The parameters of the coupling beam are: wavelength of $510$ nm, beam diameter of $1.7$ mm, power of $17$ \textmu W, and Rabi frequency of $2 \pi \times 0.97$ MHz. 
		The LO is configured to have a carrier frequency of $6.9458$ GHz and with a power of $15$ dBm. The impinging RF signals to be detected are in the vicinity of this transition frequency with a small frequency difference of $150$ kHz. We consider a bandwidth of $100$ kHz. Unless otherwise stated, the detunings are configured to zero. 
		The antenna gain of each dipole antenna is $G_{\text{Ant}} = 2.1$ dB. The LNAs used in each photodetector of the RAQ-MIMO and each RF chain of the M-MIMO have an identical gain of $G = G_{\text{LNA}} = 30$ dB. We select a typical value for the LNA noise temperature of $T = T_{\text{LNA}} = 100$ Kelvin. The quantum efficiency in the photodetection is $0.8$. 
		We consider $K = 20$, where each user is randomly distributed within a circular area that has a radius of $300$ meters. Its center is $400$ meters away from the RAQ-MIMO. Unless otherwise stated, the number of receive elements of the RAQ-MIMO is configured as $M = 300$. The large-scale fading between the RAQ-MIMO and the users is given by $-30 + 38 \log_{10} \left( 1/d \right) + F_{k}$ in dB, where $d$ is the distance between the RAQ-MIMO and the user, and $F_{k}$ is the shadow fading obeying $\mathcal{CN}(0, \sigma_{sf}^2)$. The small-scale fading obeys Rayleigh fading. The bandwidth of RAQ-MIMO is $100$ kHz. Additionally, the configurations of the conventional M-MIMO are identical to those of the RAQ-MIMO.

		\subsection{Average Achievable Rate vs. the Number of Elements}
\begin{figure*}[t!]
			\centering
			\begin{minipage}[t]{0.3\linewidth}
				\centering
				\subfloat{
					\includegraphics[width=1\textwidth]{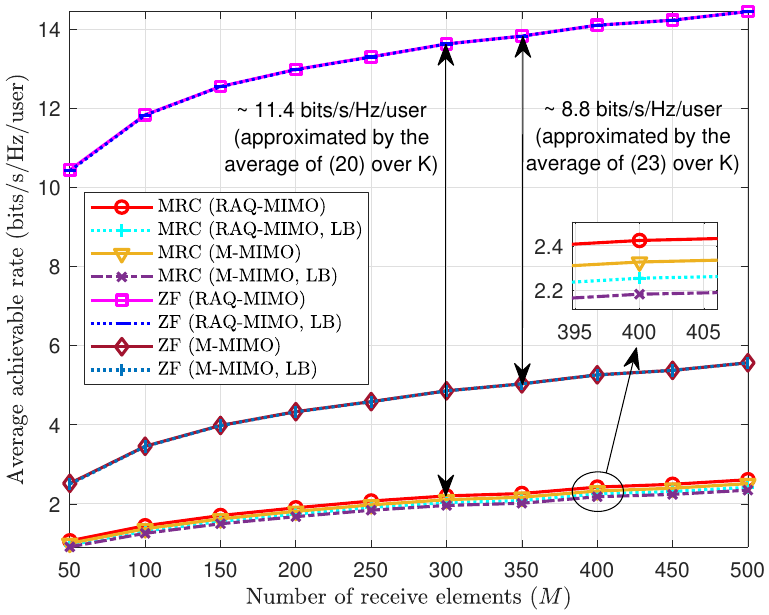}} 
				\caption{Average achievable rate vs. the number of elements.}
				\vspace{-1.2em}
				\label{fig:SEvsNumberOfElements}
			\end{minipage} \;\;
			\begin{minipage}[t]{0.3\linewidth}
				\centering
				\subfloat{
					\includegraphics[width=1\textwidth]{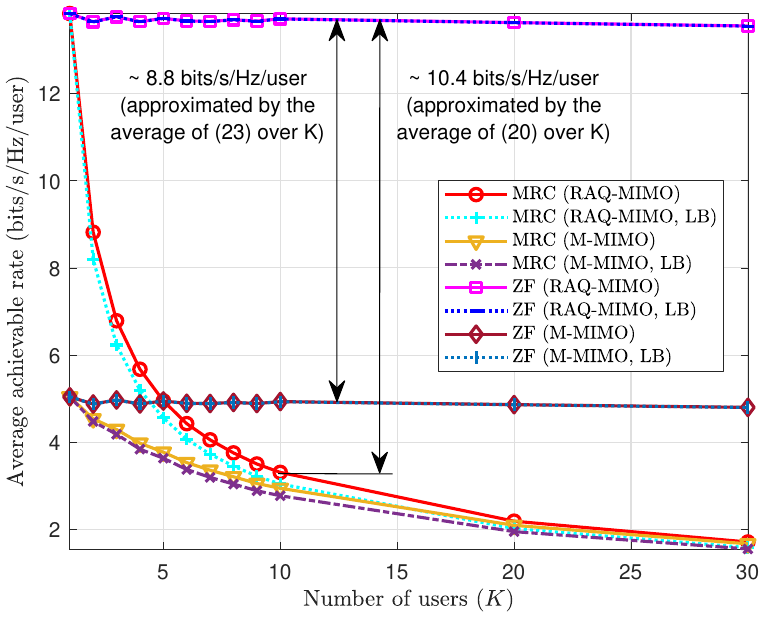}} 
				\caption{Average achievable rate vs. the number of users.}
				\vspace{-1.2em}
				\label{fig:SEvsNumberOfUsers}
			\end{minipage} \;\;
			\begin{minipage}[t]{0.3\linewidth}
				\centering
				\subfloat{
					\includegraphics[width=1\textwidth]{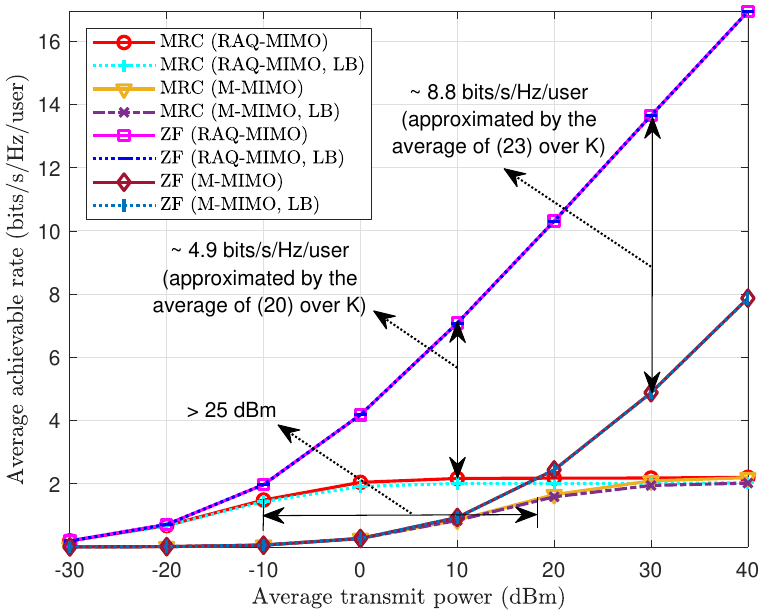}}
				\caption{Average achievable rate vs. the average transmit power.}
				\vspace{-1.2em}
				\label{fig:SEvsReceivePower}
			\end{minipage}
		\end{figure*}

		We first demonstrate the average achievable rate vs. the number of receive elements $M$ in Fig. \ref{fig:SEvsNumberOfElements}. The average transmit power is $30$ dBm and $M$ varies from $50$ to $500$. We emphasize that our theoretical results are also valid for smaller and larger $M$. 
		We observe from this figure that ZF M-MIMO outperforms MRC RAQ-MIMO, but is inferior to ZF RAQ-MIMO by an order of $\sim 8.8$ bits/s/Hz/user (approximated by the average of \eqref{eq:SE2Asmp} over $K$). The MRC curves for both RAQ-MIMO and M-MIMO are similar, which is explained in \ref{itm:C1}. When the transmit power decreases, the MRC RAQ-MIMO outperforms the MRC M-MIMO, as will be observed in \ref{subsec:SEvsTransmitPower}. 
		The ZF RAQ-MIMO achieves an extra rate of $\sim 11.4$ bits/s/Hz/user (approximated by the average of \eqref{eq:SE1} over $K$) over the MRC RAQ-MIMO. 
		Furthermore, we observe that the LBs of MRC and ZF for both RAQ-MIMO and M-MIMO are tight and reflect the true trends. 

		\subsection{Average Achievable Rate vs. the Number of Users}
		We then show the average achievable rate vs. the number of users in Fig. \ref{fig:SEvsNumberOfUsers}, where the number of RAQ-MIMO receive elements is $M = 300$ and the average transmit power is $\mathcal{P}_{s} = 30$ dBm. The number of users varies from $1$ to $30$. We can observe from the figure that the MRC curves for both RAQ-MIMO and M-MIMO gradually decrease in the average achievable rate and become identical as $K$ increases. This decrease is consistent with the trend of the expression presented in \ref{itm:C1} as $K$ increases. By contrast, the ZF curves for both RAQ-MIMO and M-MIMO are nearly flat (only decrease in slight degree), which are inline with the trend of the expression \eqref{eq:AchievableRate_ZF_LB_SC} for RAQ-MIMO and of the ZF expression \eqref{eq:MIMO_AchievableRate_MRC_LB} for M-MIMO as $K$ increases. ZF outperforms MRC for large $K$. Additionally, ZF RAQ-MIMO exhibits a significant increase in the average achievable rate compared to MRC RAQ-MIMO and ZF M-MIMO. Specifically, the average achievable rate has an extra increase of $\sim 10.4$ bits/s/Hz/user over MRC RAR-MIMO and of $\sim 8.8$ bits/s/Hz/user over ZF M-MIMO (when $K = 10$).

		\subsection{Average Achievable Rate vs. the Average Transmit Power}
		\label{subsec:SEvsTransmitPower}
		We lastly show the average achievable rate vs. the average transmit power in Fig. \ref{fig:SEvsReceivePower}, where the number of RAQ-MIMO receive elements is $M = 300$. As the average transmit power increases, the MRC curves tend to become saturated, as explained in \ref{itm:C1}, while the ZF curves continue to increase. The saturation points of MRC curves for RAQ-MIMO and M-MIMO are around $0$ dBm and $30$ dBm, respectively, which suggests that the RAQ-MIMO system allows the average transmit power of users to be much lower ($\sim 30$ dBm lower) than that of M-MIMO, as observed in Fig. \ref{fig:SEvsReceivePower}. We can see again that the MRC curves for RAQ-MIMO and M-MIMO nearly coincide upon increasing the average transmit power. ZF for RAQ-MIMO achieves an extra $\sim 8.8$ bits/s/Hz/user (approximated by the average of \eqref{eq:SE2Asmp} over $K$) over the ZF M-MIMO as the average transmit power increases.  The gap between ZF RAQ-MIMO and MRC RAQ-MIMO tends to increase, which can be estimated by the average of \eqref{eq:SE1} over $K$. For example, the gap is $\sim 4.9$ bits/s/Hz/user when the average transmit power is $10$ dBm. 
		Lastly, we observe that the LBs are tight approximations of average achievable rates.

		\section{Conclusions}
		\label{sec:Conclusions}

		In this article, we have studied the RAQ-MIMO scheme, where a receive array formed by RAQRs is applied for assisting classical uplink multiple-user communications. We have presented the functional blocks of the system and constructed the corresponding equivalent baseband signal model. The proposed scheme and the signal model pave the way for future system designs and signal processing. We have also studied the ergodic achievable rate of RAQ-MIMO under the MRC and ZF schemes. Furthermore, the tight LBs have also been derived. Based on these LBs, we have made comparisons between the two schemes for RAQ-MIMO and compared RAQ-MIMO to the conventional M-MIMO by providing closed-form expressions of the difference between their achievable rates. The results show that RAQ-MIMO allows the average transmit power of users to be $> 25$ dBm lower than that of the M-MIMO. Alternatively, an extra $\sim 8.8$ bits/s/Hz/user rate becomes achievable by ZF RAQ-MIMO. 


		\appendices
		\section{}
		\label{Appendix:LBs}
		
		For MRC, we derive $\bm{\varTheta}^{*} \bm{\varTheta} = \varrho \cos^{2} \varphi (\Omega_{l}) \bm{I}_{M}$ based on \eqref{eq:NarrowbandSampledOutputPlusNoise_MatrixForm_SC}. 
		Then $\left[ \mathbb{E} \left\{ \gamma_{k} \right\} \right]^{-1} =$ $\left[ 1 + \frac{{\varrho \cos^{2} \varphi (\Omega_{l})} {\cal P}_{s} A_e}{\sigma^{2}} \sum_{i=1, \ne k}^{K} \beta_{i} \right]$ $\frac{ 1 }{ \varrho \cos^{2} \varphi (\Omega_{l}) \frac{\varrho {\cal P}_{s} A_e}{\sigma^{2}} \mathbb{E} \left\{ \bm{h}_{k}^{*} \bm{h}_{k} \right\} } =$$\left[ 1 + \frac{{\varrho \cos^{2} \varphi (\Omega_{l})} {\cal P}_{s} A_e}{\sigma^{2}} \sum\nolimits_{i = 1, \ne k\hfill}^K \beta_{i} \right]$ $\frac{ 1 }{ \varrho \cos^{2} \varphi (\Omega_{l}) \frac{{\cal P}_{s} A_e}{\sigma^{2}} \varrho (M-1) \beta_k }$. Upon using Jensen's inequality, we get $R_{k} \ge \log_{2} \left( 1 + \mathbb{E} \left\{ \gamma_{k} \right\} \right)$. When substituting $\left[ \mathbb{E} \left\{ \gamma_{k} \right\} \right]^{-1}$ into it, we obtain \eqref{eq:AchievableRate_MRC_LB_SC}. 
		
		For ZF, we get $R_{k} \ge {\log _2}\left[ {1 + \frac{{ {\cal P}_s {A_e} {\beta _k}{\varrho}{{\cos }^2}\varphi ({\Omega _l})}}{{\sigma ^2}{\mathbb{E} \left\{ {{{\left[ {{{\left( {{{\bf{G}}^*}{\bf{G}}} \right)}^{ - 1}}} \right]}_{kk}}} \right\}}}} \right] =$ ${\log _2}\left[ {1 + \frac{{\cal P}_s {A_e}{K{\beta _k}{\varrho}{{\cos }^2}\varphi ({\Omega _l})}}{{{\sigma ^2} \mathbb{E} \left\{ {{\rm{Tr}}\left[ {{{\left( {{{\bf{G}}^*}{\bf{G}}} \right)}^{ - 1}}} \right]} \right\}}}} \right]$. Using ${\mathbb{E} \left\{ {{\rm{Tr}}\left[ {{{\left( {{{\bf{G}}^*}{\bf{G}}} \right)}^{ - 1}}} \right]} \right\}}$ $= {\frac{K}{{M - K}}}$, we obtain \eqref{eq:AchievableRate_ZF_LB_SC}.


		\balance 
		\bibliographystyle{IEEEtran}
		\bibliography{IEEEabrv,references} 

	\end{document}